\documentclass[twocolumn,final,aps,prl,showpacs,superscriptaddress,amsmath,amssymb,amsfonts,floatfix]{revtex4-1}

\usepackage[colorlinks=true,citecolor=blue,urlcolor=blue,linkcolor=blue,hyperfigures=true]{hyperref}
\usepackage{graphicx}
\usepackage{amsmath,amsfonts}
\usepackage[dvipsnames]{xcolor}
\usepackage{color}
\usepackage{soul}
\usepackage{float}

\newcommand{\orcidlink}[1]{}

\usepackage{xr}
\makeatletter
\newcommand*{\addFileDependency}[1]{
  \typeout{(#1)}
  \@addtofilelist{#1}
  \IfFileExists{#1}{}{\typeout{No file #1.}}
}
\makeatother

\newcommand*{\myexternaldocument}[1]{
    \externaldocument{#1}
    \addFileDependency{#1.tex}
    \addFileDependency{#1.aux}
}
\myexternaldocument{Supplemental}

\begin{document}

\title{Correlations turn electronic structure of finite-layer nickelates upside down}

\author{Paul Worm\,\orcidlink{0000-0003-2575-5058}}
%\thanks{These authors contributed equally}
\affiliation{Institute of Solid State Physics, TU Wien, 1040 Vienna, Austria}

\author{Liang Si\,\orcidlink{0000-0003-4709-6882}}
%\thanks{These authors contributed equally}
\email{liang.si@ifp.tuwien.ac.at}
\affiliation{Institute of Solid State Physics, TU Wien, 1040 Vienna, Austria}

\author{Motoharu Kitatani\,\orcidlink{0000-0003-0746-6455}}
\affiliation{RIKEN Center for Emergent Matter Sciences (CEMS), Wako, Saitama, 351-0198, Japan}

\author{Ryotaro Arita\,\orcidlink{0000-0001-5725-072X}}
\affiliation{RIKEN Center for Emergent Matter Sciences (CEMS), Wako, Saitama, 351-0198, Japan}
\affiliation{Department of Applied Physics, The University of Tokyo, Hongo, Tokyo, 113-8656, Japan}

\author{Jan M. Tomczak\,\orcidlink{0000-0003-1581-8799}}
\affiliation{Institute of Solid State Physics, TU Wien, 1040 Vienna, Austria}

\author{Karsten Held\,\orcidlink{0000-0001-5984-8549}}
\email{held@ifp.tuwien.ac.at}
\affiliation{Institute of Solid State Physics, TU Wien, 1040 Vienna, Austria}

\date{\today}

\begin{abstract}
Motivated by the recent discovery of superconductivity in the pentalayer  nickelate Nd$_6$Ni$_5$O$_{12}$ [Nature Materials 10.1038], we calculate its electronic structure and superconducting critical temperature.
We find that electronic correlations are essential for pushing Nd$_6$Ni$_5$O$_{12}$ into the superconducting doping range as they shift the electron pockets above the Fermi energy. 
As a consequence, Nd$_6$Ni$_5$O$_{12}$  can be described with a single $d_{x^2-y^2}$ orbital per Ni. Instead, for the bilayer nickelate Nd$_3$Ni$_2$O$_6$ we find correlations to drive the system into a three-orbital regime also involving the Ni $d_{xz,yz}$ states. 
We suggest, however, that single-orbital physics with optimal doping can be restored by substituting 60\% of the trivalent Nd or La by tetravalent Zr.
\end{abstract}

\maketitle

Even 35 years after the discovery of high temperature ($T_c$) superconductivity in cuprates \cite{Bednorz1986}, understanding the microscopic mechanism and identifying the 
necessary ingredients for a minimal model remains the arguably biggest challenge  of solid state theory. The recent synthesis of nickelate superconductors   Sr(Ca)$_x$Nd(La,Pr)$_{1-x}$NiO$_2$ \cite{li2019superconductivity,zeng2020,Osada2020,Osada2021,Zeng2021} provides
a new perspective and has awoken
new hope for this quest, not least because nickelates are very similar, but, at the same time, also very distinct from cuprates. In conjuncture with the cuprates they are hence  ideally suited to distinguish the essential from the incidental for high-$T_c$ superconductivity. That is, Nd(La,Pr)NiO$_2$ shares the same infinite layer structure and formal 3$d^9$ valence with the simple cuprate superconductor  CaCuO$_2$ \cite{DiCastro2012,DiCastro2015}. However, besides the $d_{x^2-y^2}$ band, additional electron pockets  are present at the Nd(La,Pr)NiO$_2$  Fermi surface, as identified in electronic structure calculations \cite{Botana2019,Hirofumi2019,jiang2019electronic,Motoaki2019,hu2019twoband,Wu2019,Nomura2019,Zhang2019,Jiang2019,Werner2019,Si2019,Kitatani2020} and evidenced by a negative Hall coefficient \cite{li2019superconductivity,zeng2020}.

Since infinite layer cuprates are not the best superconductors, finding superconductivity in a finite layer nickelate, the undoped pentalayer  Nd$_{6}$Ni$_{5}$O$_{12}$  \cite{pan2021}, must be considered a breakthrough for nickelate superconductivity. In contrast to infinite layer nickelates, it has  a positive Hall coefficient \cite{pan2021}, although density functional theory (DFT) predicts intriguingly large electron pockets. The trilayer nickelate Nd$_{4}$Ni$_{3}$O$_{8}$ is, on the other hand, not superconducting  \cite{pan2021}. This finding naturally leads to the question which finite layer nickelates can be made superconducting and what doping is required to this end.

\begin{figure*}[!ht]
\begin{minipage}{0.75\textwidth} \hfill
\includegraphics[width=\textwidth]{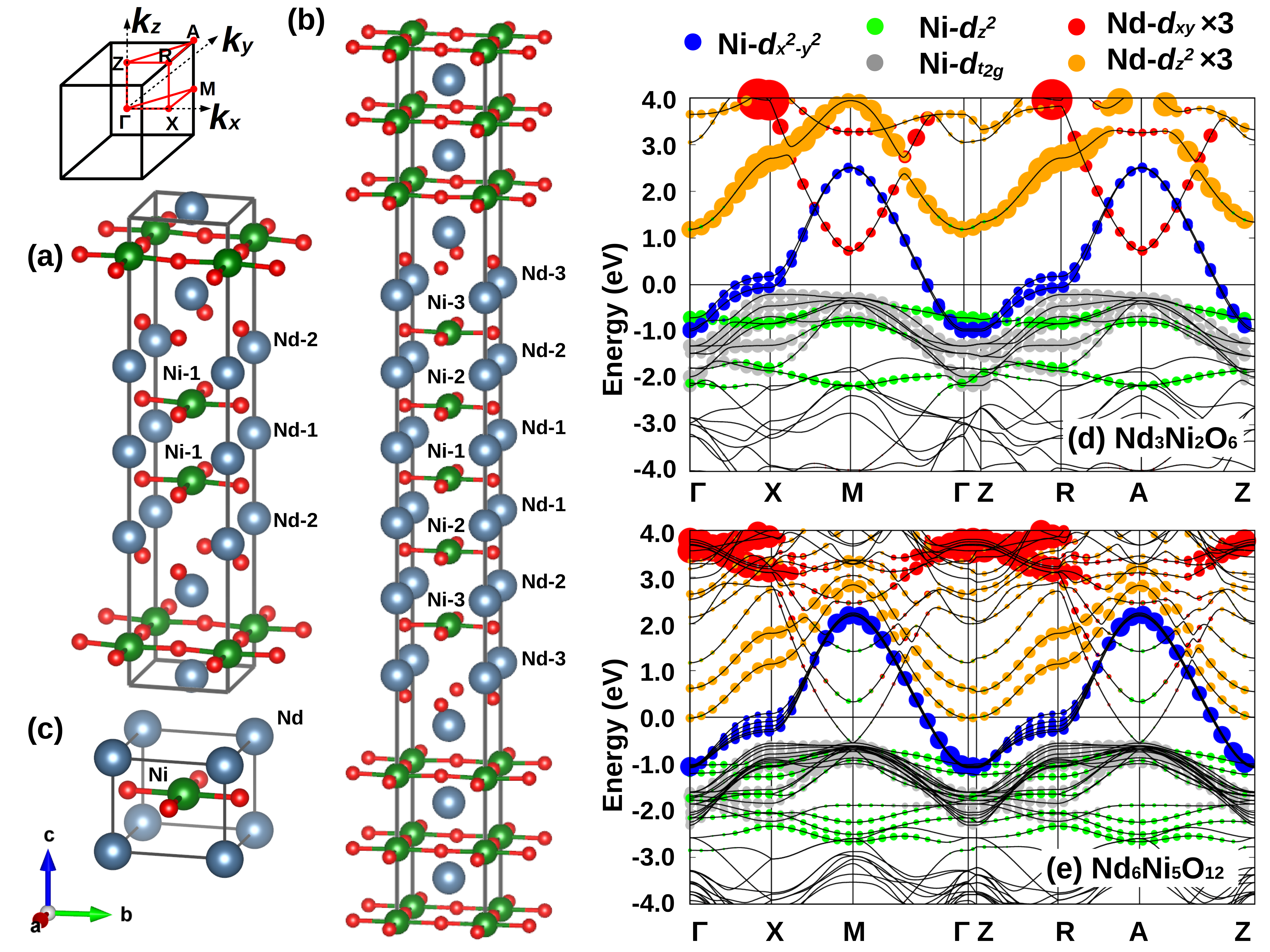}\hfill
\end{minipage}\hfill
\begin{minipage}{0.21\textwidth}
\vspace{.0cm}
\caption{Crystal structure of the reduced Ruddlesden-Popper  perovskites Nd$_{n+1}$Ni$_n$O$_{2n+2}$: (a) Nd$_3$Ni$_2$O$_6$ ($n$=$2$), (b)  Nd$_6$Ni$_5$O$_{12}$ ($n$=$5$) and (c) NdNiO$_2$ ($n$=$\infty$). From the center to the interface layer, the inequivalent Nd (Ni) atoms are labeled as Nd-1, 2, 3 (Ni-1, 2, 3).  Nd, Ni and O atoms are shown as blue, green and red balls, respectively. The top-left panel shows the Brillouin zone and the notation for the high-symmetry momenta. 
DFT band structure and orbital character for (d) Nd$_3$Ni$_2$O$_6$ and (e) Nd$_6$Ni$_5$O$_{12}$ along a high symmetry path through the Brillouin zone.}
\label{Fig1}
\end{minipage}
\end{figure*}

In this letter, we analyze the  superconducting pentalayer Nd$_{6}$Ni$_{5}$O$_{12}$ and  the arguably simplest finite-layer compound,
bilayer Nd$_{3}$Ni$_{2}$O$_{6}$, by state-of-the-art  density functional theory (DFT) plus dynamical mean field theory (DMFT) \cite{RevModPhys.68.13,Anisimov1997,Lichtenstein1998,held2007electronic}. 
We show that local DMFT correlations push the Nd pockets (that exist in DFT \cite{pan2021}) above the Fermi level, thereby leaving only one correlated $d_{x^2-y^2}$ orbital per layer to form the Fermi surface---reminiscent of cuprate superconductors. Furthermore, we compare the effective $d_{x^2-y^2}$ model of this pentalayer system to the infinite layer case and find a stunningly similar tight-binding parameters and effective masses.
The main differences between the pentalayer and infinite layer system are two-fold: (i) Nd pockets are only present in the infinite layer compound if electronic correlations are properly taken into account. (ii) The five layer structure has, every five layers, a fluorite blocking slab that essentially cuts off the (weak) interlayer hopping every five layers. 
Hence, our DFT+DMFT calculations strongly suggest a single-orbital per Ni Hubbard model as the minimal model. Neglecting the weak inter-layer hopping results in a 2D one-band Hubbard model description, which we further study by including
non-local correlations within the
dynamical vertex approximation (D$\Gamma$A) \cite{Toschi2007,Rohringer2018}. We find that spin fluctuations are dominant and give rise to superconductivity with a $T_c$ comparable to experiment.

As for the undoped bilayer Nd$_3$Ni$_2$O$_6$ with formal Ni valence $d^{8.5}$,
we show that local DMFT correlations drive a charge transfer from the Ni $d_{xz/yz}$ to the $d_{x^2-y^2}$ orbitals, resulting in a {\it per se} multi-orbital system with a three-dimensional electronic structure. Remarkably, however, we find that doping to a nominal valence of $d^{8.8}$, the $d_{xz/yz}$ orbitals become fully populated again and the desirable one-orbital physics is recovered:
The thus doped bilayer system again shows properties similar to the infinite and pentalayer superconductors. As a dopant we propose Zr, specifically La$_{2+\delta}$Zr$_{1-\delta}$Ni$_{2}$O$_{6}$ with $\delta = 0.4$.

\emph{Methods.}
Structural relaxations for lattice constants and atomic positions are performed with the \textsc{vasp} code \cite{PhysRevB.48.13115}, while the \textsc{wien2k} package is used  for the DFT electronic structures \cite{blaha2001wien2k}. For both, the PBE \cite{PhysRevLett.77.3865} and PBESol \cite{PhysRevLett.100.136406} versions of the generalized gradient approximation (GGA) are employed on a dense momentum grid with 3000 $k$-points.

For the DMFT calculations, the \textsc{wien2k} bands around the Fermi level are projected onto maximally localized Wannier functions \cite{PhysRev.52.191,RevModPhys.84.1419} using the \textsc{wien2wannier} interface \cite{mostofi2008wannier90,kunevs2010wien2wannier} and supplemented by local Kanamori interaction, taking the fully localized limit \cite{Anisimov1991} as double counting. For Nd$_3$Ni$_2$O$_{6}$ a projection onto only the Ni-3$d$ orbitals is sufficient, since the Nd-5$d$ bands are here well separated and above the Fermi level, as shown in Fig.~\ref{Fig1}(d) [discussed below]. Contrary, for Nd$_6$Ni$_5$O$_{12}$ a projection onto both, the Ni-3$d$ and Nd-5$d$ shell, is necessary to account for the electron pocket around the $M$- and $A$-points, that are present in DFT and stem from Nd-5$d$ bands. Additionally, at the $\Gamma$-point there is a Nd-$d_{z^2}$ band that barely touches the Fermi level  E$_F$. 
Expecting only minor variations with the number of nickel-layers, we use the same interaction parameters as were obtained for LaNiO$_2$\cite{Si2019}:
An  average inter-orbital interaction $U'= 3.10\,$eV ($2.00\,$eV) and Hund's exchange  $J=0.65\,$eV  ($0.25\,$eV) for Ni (Nd). 
The intra-orbital Hubbard interaction follows as $U=U'+2J$. These interaction parameters are close to those of  previous studies \cite{PhysRevB.73.155112,PhysRevLett.115.236802} for 3$d$ oxides. The resulting Hamiltonian is then solved at room temperature (300\,K) in DMFT using continuous-time quantum Monte Carlo simulations in the hybridization expansions \cite{RevModPhys.83.349} implemented in \textsc{w2dynamics} \cite{PhysRevB.86.155158,wallerberger2019w2dynamics}. The maximum entropy method \cite{PhysRevB.44.6011,PhysRevB.57.10287,Kaufmann2021} is employed for an analytic continuation of the spectra.

As we will see below, a one-band per Ni site description based on the  Ni-$d_{x^2-y^2}$ orbital is possible at low energies.
Hence, we perform a further Wannier projection onto this $d_{x^2-y^2}$ orbital 
only [see supplemental material (SM) \cite{SM} Section S.2 for further information]. We supplement this effective tight-binding one-band model with an on-site interaction value of $U$=8$t$, motivated by cRPA for the single-band projection in agreement with previous studies \cite{Kitatani2020}. This one-band model for five isolated layers
can then be treated by more involved many-body techniques. We use the dynamical vertex approximation (D$\Gamma$A) \cite{Toschi2007,Rohringer2018}, employing an additional, admittedly crude, approximation: neglecting the interlayer hopping. The thus obtained estimate for the critical temperature $T_c$ is obtained along the lines of \cite{Kitatani2019}.

%--------------------------------------------------------------------------------
\emph{DFT crystal and electronic structure.}
%--------------------------------------------------------------------------------
We consider the series Nd$_{n+1}$Ni$_n$O$_{2n+2}$ which can be synthesized from the Ruddlesden-Popper perovskite parent compound by oxygen reduction. 
Specifically, we concentrate on  Nd$_3$Ni$_2$O$_6$ ($n$=$2$), Nd$_6$Ni$_5$O$_{12}$ ($n$=$5$, experimentally realized in \cite{pan2021})  and NdNiO$_2$ ($n$=$\infty$, experimentally realized in \cite{li2019superconductivity,zeng2020,Li2020Dome}). Their crystal structures are shown in
Fig.~\ref{Fig1}.
Note the interface NdO-NdO layers between the $n$ NiO$_2$ layers are transformed to fluorite-like Nd-O$_2$-Nd blocking slabs, which we find energetically favorable in DFT+$U$  compared to the octahedral rock salt interface by 2.019\,eV and 2.242\,eV 
[per supercell as in Fig.~\ref{Fig1}(a, b)]
 for Nd$_3$Ni$_2$O$_6$ and Nd$_6$Ni$_5$O$_{12}$, respectively.

An intriguing question is: how do the electronic bands change with the number of layers $n$? In this respect the presence or absence of pockets derived from the Nd-5$d$ bands is of particular importance.
DFT studies of NdNiO$_2$ \cite{Nomura2019,Zhang2019,Dasgupta2020,Hirofumi2019,gu2020substantial}
show an electron pocket at the $A$- and $\Gamma$-point originating from the  Nd-5$d$ orbitals.
Electronic correlations and Sr-doping may push the $\Gamma$-pocket above E$_F$. In contrast, the $A$-pocket remains present in the superconducting doping region and acts 
as an electron reservoir \cite{Kitatani2020} that (self)dopes the Ni 3$d$ bands away from the nominal 3$d^9$ configuration. 
In contrast,  Fig.~\ref{Fig1}(d) for
Nd$_3$Ni$_2$O$_6$ shows no electron pockets; the Nd-bands are clearly above E$_F$ and well separated from the Ni bands. This indicates the hybridization between Nd and Ni is weaker in Nd$_3$Ni$_2$O$_6$, i.e., the fluorite interface makes both NiO$_2$ and Nd-O$_2$-Nd more 2-dimensional (2D). For Nd$_6$Ni$_5$O$_{12}$ in Fig.~\ref{Fig1}(e), the DFT 
bandstructure has again a tube-like pocket including momenta $A$ and $M$ (see SM \cite{SM} for a plot) and a band touching E$_F$ at 
$\Gamma$.
The latter is dominantly composed of Nd-$d_{z^2}$, similar to the $\Gamma$-pocket in NdNiO$_2$ \cite{Nomura2019,Hirofumi2019,gu2020substantial}. The tube around $A$- and $M$ is composed of Ni-$d_{z^2}$ and Nd-$d_{xy}$ character.

To further investigate prospective similarity between infinite and finite layers, we list the electronic hopping terms from a  Wannier projection in Table~\ref{table:hopping}. Infinite layer NdNiO$_2$ and pentalayer Nd$_6$Ni$_5$O$_{12}$ have almost identical in-plane hopping parameters. Also the out-of-plane $t_z$ hopping is almost the same within the five layers of the pentalayer nickelate, while it is practically zero across the fluorite spacing layer. 
Additionally both systems have very similar ratios of $t'/t$ and $t''/t$. The bilayer Nd$_3$Ni$_2$O$_6$ has  $\sim$5\% $t$ larger $t$ which  can be attributed to the increased two-dimensionality caused by the fluorite interface, which cuts of $z$-direction hopping after every second layer, rather than every fifth in the pentalayer.

While the hopping is similar, the  filling of the  $d_{x^2-y^2}$-band,  a key  factor for superconductivity, differs between the compounds. Nd$_3$Ni$_2$O$_6$ hosts a nominal $d^{8.5}$ configuration which is far away from the optimal  hole doping $\delta\lesssim$0.2 ($d^{9-\delta}$).  Also the doping of  Nd$_6$Ni$_5$O$_{12}$, as predicted by DFT, is not within the superconducting region. The $A$-$M$-pocket takes electrons away, leading to $\delta\sim0.24$ holes (see SM \cite{SM} Section~S.3 and Table~S.I) in the effective Ni-$d_{x^2-y^2}$ band crossing E$_F$, instead of the formal valency $\delta=0.2$ ($d^{8.8}$).
As we will see next, electronic correlations, however, shift the tube-like pocket around  $A$- and $M$ above the Fermi energy so that superconducting doping is indeed achieved.

\begin{table}
\caption{Major tight-binding hopping parameters (in units of eV) for the  Ni-$d_{x^2-y^2}$ orbital as obtained from a single-band Wannier projection; $t$, $t'$, $t''$ and $t_z$ indicate 1st [hopping along the direction $R=$(100)], 2nd [$R$=(010)], 3rd nearest neighbor [$R$=(200)] and z-direction [$R$=(001)] hopping, respectively. For Nd$_6$Ni$_5$O$_{12}$, Ni-1 is the central layer, Ni-3 denotes the two equivalent interface layers, and Ni-2 the two layers in-between, see Fig.~\ref{Fig1}(b).}
\label{table:hopping}
\begin{tabular}{c|c|c|c|c|c|c}
\hline
System             & $t$      & $t'$    & $t''$    & $t_z$  & $t'/t$   & $t''/t$  \\ \hline \hline
NdNiO$_2$          & -0.395 & 0.095 & -0.047 & -0.034 & -0.242 & 0.119 \\ \hline \hline
Nd$_3$Ni$_2$O$_6$  & -0.414 & 0.092 & -0.055 & -0.055 & -0.223 & 0.132 \\ \hline \hline
 Nd$_6$Ni$_5$O$_{12}$: Ni-1    & -0.395 & 0.098 & -0.050 &       & -0.249 & 0.127 \\ \cline{1-4}\cline{6-7} 
 &       &       &      & -0.031 &      &     \\  
Nd$_6$Ni$_5$O$_{12}$: Ni-2        & -0.392 & 0.097 & -0.050 &     \hrulefill    & -0.249 & 0.127 \\ \cline{1-4}\cline{6-7}
 &        &       &        & -0.026 &        &       \\ 
 Nd$_6$Ni$_5$O$_{12}$: Ni-3 & -0.398 & 0.097 & -0.049 &         & -0.245 & 0.122 \\ \hline
\end{tabular}
\end{table}

\begin{figure}[tb]
\includegraphics[width=\columnwidth]{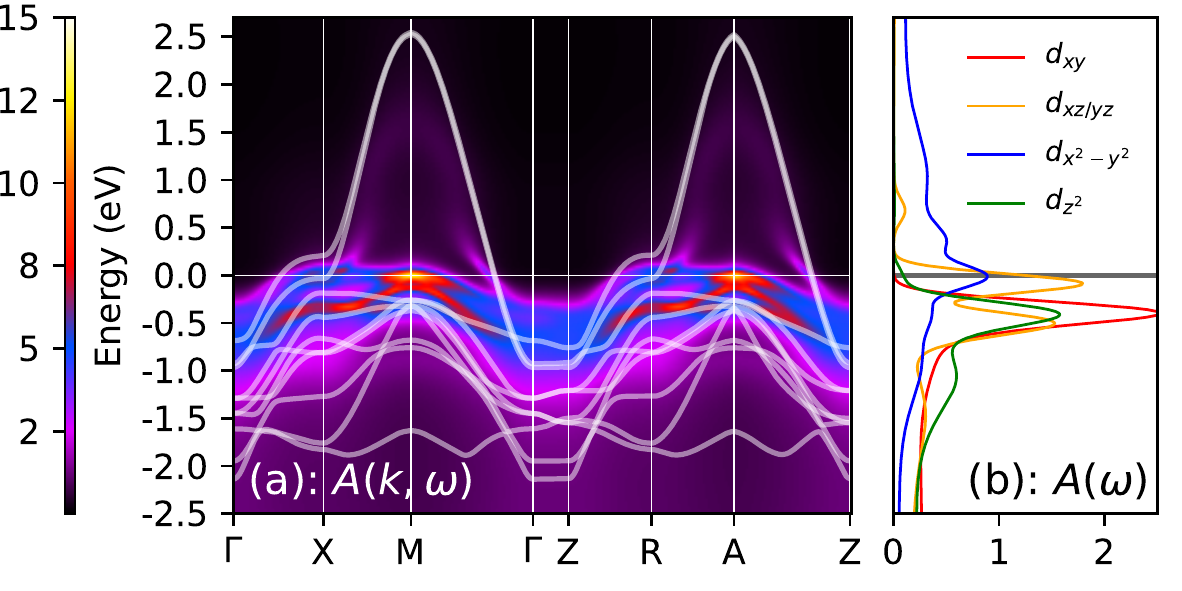}
\caption{(a) DMFT $k$-resolved spectral function $A(k,\omega)$ for Nd$_3$Ni$_2$O$_6$. White lines are the DFT Wannier bands. (b) Corresponding $k$-integrated spectral function $A(\omega)$. Local correlations push the $d_{xz/yz}$ orbitals up in energy, which now contribute to the Fermi-surface at the $M$-point (cf.\ SM Fig.~S7).}
\label{Fig3}
\end{figure}

\begin{figure}[tb]
\includegraphics[width=\columnwidth]{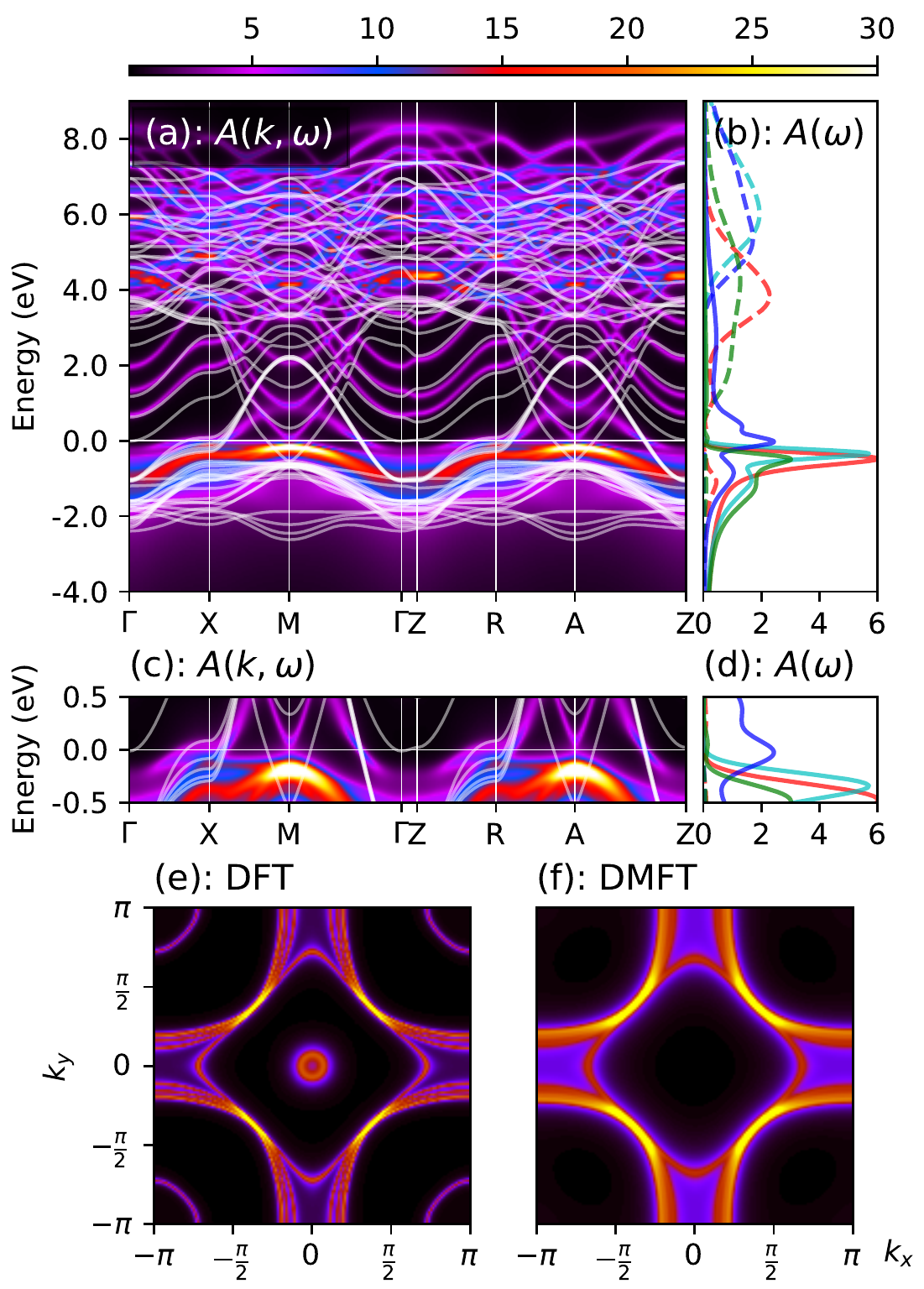}
\caption{DMFT $k$-revolved spectral functions $A(k,\omega)$  (a) and $k$-integrated A($\omega$) (b) of Nd$_6$Ni$_5$O$_{12}$; (c) and (d) are a zoom-in  around E$_F$=0; solid lines in (a, c) are the DFT Wannier-bands. Fermi surface  for (e)  DFT and (f)  DMFT. The colorbar at the top is for all plots. The DMFT Fermi surface has been multiplied by two for better visibility.}
\label{Fig4}
\end{figure}

%--------------------------------------------------------------------------------
\emph{DMFT results.}
%--------------------------------------------------------------------------------
For both, Nd$_3$Ni$_2$O$_6$ and  Nd$_6$Ni$_5$O$_{12}$, electronic correlations dramatically change the electronic structure, see Figs.~\ref{Fig3} and \ref{Fig4}. %In the case of 
In Nd$_3$Ni$_2$O$_6$, Hund's exchange favors a more equal occupation of %the
orbitals since this way a spin-1 can be formed and also the Coulomb repulsion is lower ($U'=U-2J$ instead of $U$). 
As a result, local DMFT correlations drive an electron transfer from the $d_{xz/yz}$ into the $d_{x^2-y^2}$ orbital, see SM \cite{SM} Table~S.I. The now partially depopulated $d_{xz/yz}$ orbitals are pushed up in energy,  both $d_{x^2-y^2}$ and $d_{xz/yz}$ form quasiparticle peaks in Fig.~\ref{Fig3} and contribute to the Fermi surface (see SM \cite{SM} Section~S.5-S.6). This multi-orbital, three-dimensional electronic structure is, to the best of our knowledge, not favorable for superconductivity.

For the pentalayer, %system 
Nd$_6$Ni$_5$O$_{12}$, electronic correlations have another game-changing effect, see Fig.~\ref{Fig4}: 
The Nd-pocket around $A$- and $M$-momentum is pushed above the Fermi level,  which only leaves the Ni $d_{x^2-y^2}$ contributing to the Fermi-surface (cf.\ SM \cite{SM} Section~S.5-S.6). We thus have an ideal one-orbital system that is quite two-dimensional and prone to strong antiferromagnetic fluctuations, as the predominant
Ni $d_{x^2-y^2}$ bands now hosts $\delta =0.19$, $0.21$, and $0.20$ holes in the three inequivalent layers, compared to $\delta =0.24$, $0.27$, and $0.26$ in DFT. 
As seen below, this pushes Nd$_6$Ni$_5$O$_{12}$ into the superconducting doping regime. Electronic correlations further  enhance the effective mass to $m^*/m_b \sim 2.5$ (see SM \cite{SM} Section~S.4. As for the hopping parameters, this is almost the same as in superconducting Sr$_{0.2}$Nd$_{0.8}$NiO$_2$ \cite{Kitatani2020}.
Altogether, the DFT+DMFT results hence show a striking similarity of the pentalayer and the infinite layer system, with the only noteworthy difference being the Nd $A$-pocket in NdNiO$_2$. However, this pocket has been argued to only be a passive charge reservoir for superconductivity \cite{Kitatani2020}.

This brings us back to the bilayer compound Nd$_3$Ni$_2$O$_6$ and the question: can we tune its electronic structure to one that is favorable for superconductivity? Indeed we believe one can do so through chemical doping/substitution. Here we consider
La$_{2+x}$Zr$_{1-x}$Ni$_2$O$_6$ which has a nominal valence configuration 3$d^{9-x/2}$. At $x=0.4$ no electron pockets are present within DFT+DMFT anymore, and the Fermi surface is composed only of a single Ni $d_{x^2-y^2}$ orbital (see SM~\cite{SM} Section~S.7), hosting $\delta=\frac{x}{2}=0.2$ holes, which is quite ideal for superconductivity.

\begin{figure}[tb]
\includegraphics[width=\columnwidth]{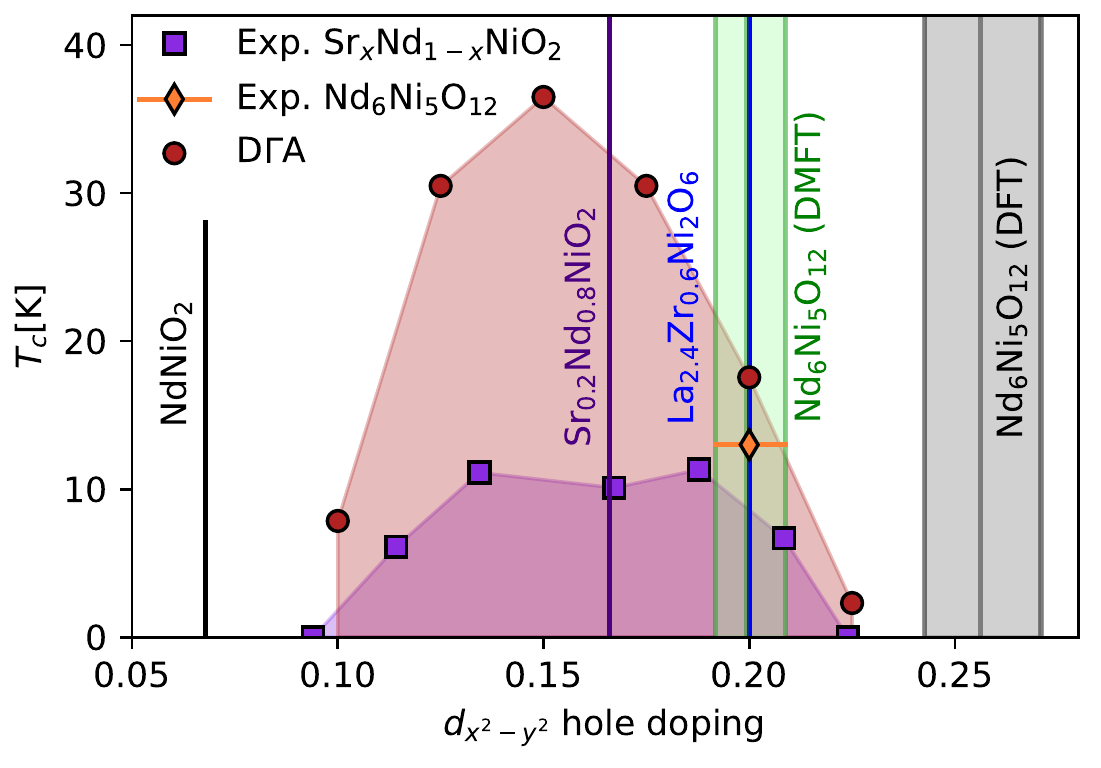}
\caption{Phase diagram $T_c$ vs. hole doping of the Ni $d_{x^2-y^2}$ band as calculated by  D$\Gamma$A. For Nd$_6$Ni$_5$O$_{12}$, the 
doping of the inequivalent layers is marked by three green (DMFT) 
and gray lines (DFT) and the experimental $T_c$ (diamond) is from \cite{pan2021}. For Sr-doped  NdNiO$_2$, experimental data are from
\cite{zeng2020}.}
\label{Fig5}
\end{figure}

%--------------------------------------------------------------------------------
\emph{Estimating $T_c$ with D$\Gamma$A.}
The subsequent single-orbital DMFT calculation for Nd$_6$Ni$_5$O$_{12}$
agrees quite well with the  fully fledged calculation with all Ni 3$d$ orbitals, see SM~\cite{SM} Section~S.5-S.6.
The same is true for the effective masses $m^*/m_b$ 
(compare SM \cite{SM} Table~S.I and S.III),  justifying altogether the $d_{x^2-y^2}$-only projection.

Since the inter-layer hoppings are equally small as in the infinite layer system, we now treat the layers as independent as done previously for NdNiO$_2$ \cite{Kitatani2020}. 
This allows us to calculate $T_c$ within D$\Gamma$A as a function of the $d_{x^2-y^2}$ orbital-filling
for the pentalayer nickelate. The $D\Gamma A$ solution is displayed as the red-shaded region in Fig.\ref{Fig5}, experiment  \cite{pan2021}  as a diamond, while the experimental phase diagram for NdNiO$_2$ is the violet-shaded region taken from \cite{zeng2020} and translated to the $d_{x^2-y^2}$ filling as detailed in the SM~\cite{SM}. 
As we can see, the doping of the parent compound Nd$_6$Ni$_5$O$_{12}$ is indeed already in the optimal range for superconductivity.
This was not the case without correlations: DFT (gray lines in Fig.\ref{Fig5}) would predict a doping outside of the superconducting dome, due to the presence of the tube around $A$- and $M$.
Also La$_{2.4}$Zr$_{0.6}$Ni$_2$O$_6$ is within the superconducting doping range (blue line), but it has slightly different hopping parameters, in particular a larger $t$ which should result in a slightly larger $T_c$ as (i) $t$  sets the unit of energy and (ii) according to D$\Gamma$A, a smaller $U/t$ is favorable for $T_c$.

%--------------------------------------------------------------------------------
\emph{Conclusion.}
We have seen that electronic correlations completely alter the electronic structure of finite layer nickelates, even beyond an effective mass renormalization of the bands. In the case of the bilayer nickelate Nd$_3$Ni$_2$O$_6$, DFT predicts a  single  Ni $d_{x^2-y^2}$ band to cross the Fermi energy that accommodates $\sim$0.5 holes per Ni site (leading to 3$d^{8.5}$). For such a large hole doping, electronic correlations push the $d_{xy/yz}$ bands across the Fermi level. In the case of the pentalayer nickelate Nd$_6$Ni$_5$O$_{12}$, DFT predicts an electron pocket around the $A$ and $M$ momenta. This tube-shaped DFT pocket would dope the  Ni $d_{x^2-y^2}$ band to $0.24-0.27$ holes for the three inequivalent layers, beyond the doping range where superconductivity is expected. However, electronic correlations push the electron pocket above the Fermi energy so that there are $0.2$ holes per Ni with only a tiny variation of $~0.02$ holes between the layers. This way the Ni $d_{x^2-y^2}$ band is placed in the optimal superconducting doping range; it also has virtually the same hopping terms as  NdNiO$_2$. Hence electronic correlations are key to explain the experimental observation of superconductivity in Nd$_6$Ni$_5$O$_{12}$ \cite{pan2021}. 

While for NdNiO$_2$ the $A$-pocket is possibly only a passive bystander for superconductivity \cite{Kitatani2020}, it will certainly contribute to the Hall coefficient. The very different Hall coefficient for $n=\infty$ and $n=5$  can thus be explained by the presence (NdNiO$_2$) or absence (Nd$_6$Ni$_5$O$_{12}$) of these electron pockets.
Given the layer-dependent doping of Nd$_6$Ni$_5$O$_{12}$, it is tempting to attribute, at least part of, the unusually broad superconducting transition to this  doping variation.

Our phase diagram in Fig.~\ref{Fig5} suggests that somewhat less holes in the $d_{x^2-y^2}$ band is  still within the superconducting dome.
Without electron pockets,  Nd$_{n+1}$Ni$_n$O$_{2n+2}$ has a hole doping $\delta=1/n$. Hence up to about $n=10$, the parent compound is in the superconducing doping range.
For $n<5$,  we have instead too many holes;
doping with a tetravalent cation as, e.g., in La$_{2.4}$Zr$_{0.6}$Ni$_2$O$_6$  is needed.

\subsection{Acknowledgments}

\begin{acknowledgments}
We thank J.~Kaufmann  for helpful comments and discussions. 
%L.\,S. thanks the initial funds from Northwest University (China).
This work was supported by the Austrian Science Fund (FWF) through projects P~30819 and P~32044, by the Japan Society for the Promotion of Science (JSPS) KAKENHI Grant Number 19H05825,JP20K22342 and JP21K13887.
Calculations have been done on the Vienna Scientific Cluster (VSC).
\end{acknowledgments}

%\bibliography{full}
%\bibliographystyle{apsrev4-1}

%

%\input{main.bbl}

\clearpage

\setcounter{equation}{0}
\setcounter{figure}{0}

\onecolumngrid

{\large \textbf{Supplemental Material to Correlations turn electronic structure of finite-layer nickelates upside down}}

%

%

%\begin{abstract}
\large{ This Supplementary Material presents additional information as well as DFT and DMFT results supporting our conclusions in the main text: Section \ref{sec:dmft_details} provides computational details regarding DMFT and analytic continuation, while Section \ref{sec:dft_wannier} provides details on the Wannier projection,and Section~\ref{sec:orbital_occ} on  the calculation of the occupation of the {\em effective } Ni 3$d_{x^2-y^2}$  orbital. Section~\ref{sec:effmass} compares the effective mass for different nickelates and different Wannier projections. Section~\ref{sec:d_only_projection} supplements the information of the main text by the DMFT spectral functions for other nickelates and other Wannier function projections. Section~\ref{sec:fs} shows the corresponding Fermi surfaces. Finally, Section~\ref{sec:La2ZrNi2O6} presents
results for La$_{2+\delta}$Zr$_{1-\delta}$Ni$_2$O$_6$ at $\delta=0$ and $0.4$. }
%\end{abstract}

%\maketitle

% ----------------------------------------------------------
\section{1. Computational details on DMFT and analytic continuation} 
\label{sec:dmft_details}
% ----------------------------------------------------------
As noted in the method section of the main text, all DMFT calculations were performed using the w2dynamics code \cite{wallerberger2019w2dynamics}, which solves the Anderson impurity model using  continuous-time  quantum Monte Carlo simulations in the hybridization expansions \cite{RevModPhys.83.349}. We construct a separate impurity problem for each Nd and Ni site, which are connected via the DMFT bath (Dyson equation). After converging the DMFT iteration, we perform an additional high-statistic run to improve our analytical continuation. For this run we perform of the order $~5 \cdot 10^9$ and $~2 \cdot 10^8$ measurements for the $d_{x^2-y^2}$-only and full-$d$ projection, respectively. 
 
Analytic continuation is performed by the Maximum entropy method \cite{PhysRevB.44.6011,PhysRevB.57.10287} using the ana-cont package \cite{Kaufmann2021}. To determine the hyperparamter $\alpha$, which is one of the most crucial parameters in the fitting procedure, we employ the "chi2kink" \cite{Kraberger2017,Bergeron2016} method. Additionally, we employ a "preblur" width of $0.1$ to avoid unphysical artifacts around the Fermi energy.

% ----------------------------------------------------------
\section{2. Wannier projections} 
\label{sec:dft_wannier}
% ----------------------------------------------------------

\begin{figure}[ht]
\includegraphics[width=\textwidth]{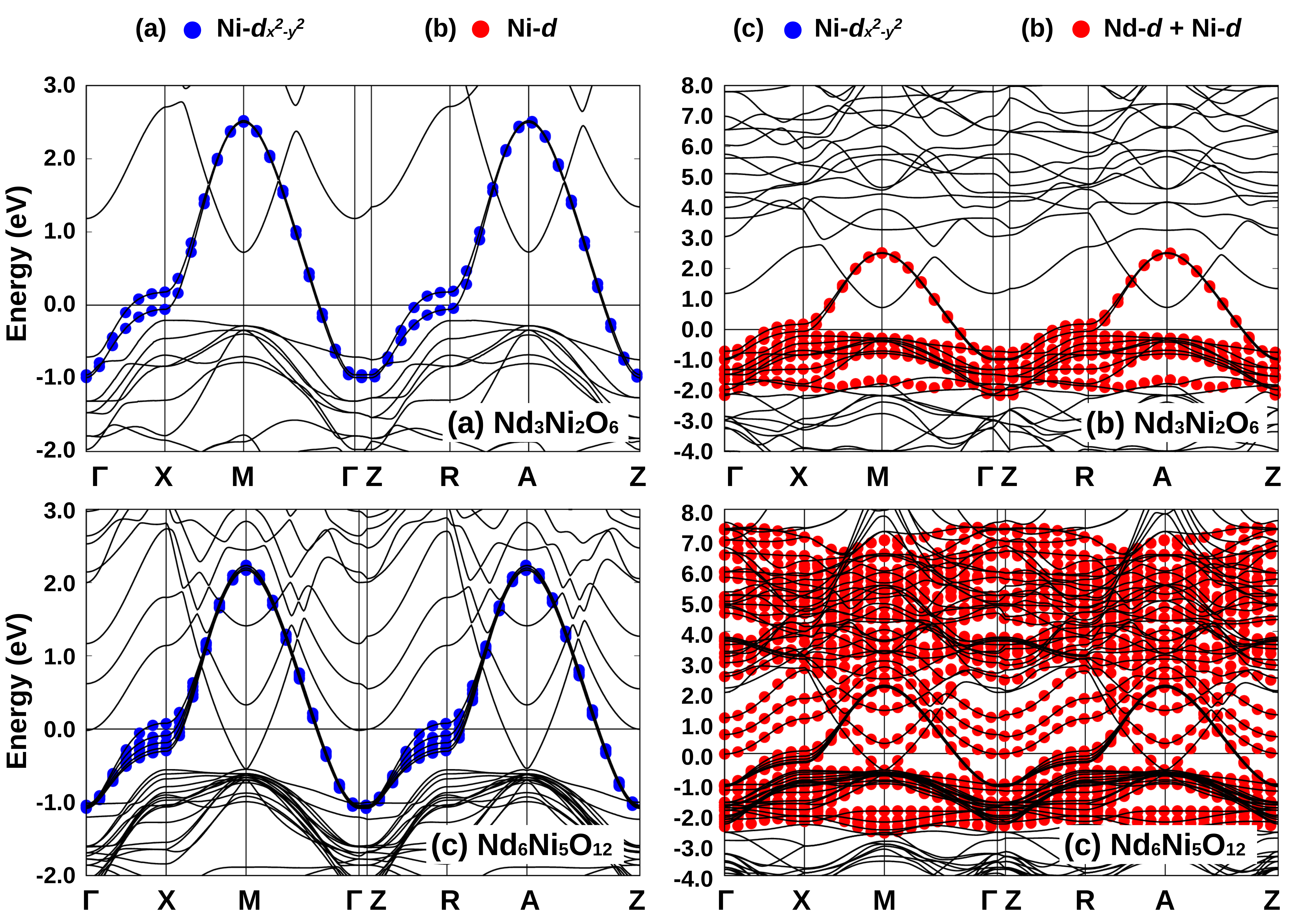}
\caption{DFT bands (black lines) and Wannier projection bands for the full $d$-shell (red, right) and the $d_{x^2-y^2}$ band only (blue, left), Nd$_{3}$Ni$_{2}$O$_{6}$ (a, b) and Nd$_6$Ni$_5$O$_{12}$ (c, d).}
\label{SFig1_wannier}
\end{figure}

In this section of the supplementary material we provide additional information concerning the Wannier projection of the main text. Specifically, we use the following Wannier projections: (i) projection of the full $d$-shell for Ni and Nd (Nd$_6$Ni$_5$O$_{12}$) and Ni-$d$ only (Nd$_{3}$Ni$_{2}$O$_{6}$); (ii) projection onto the Ni 3$d_{x^2-y^2}$ bands only. We use the $d_{x^2-y^2}$-only projection as basis for our single-band low-energy effective Hamiltonian. Such a Ni $d_{x^2-y^2}$-only projection reproduces the corresponding DFT bands well (Fig.~\ref{SFig1_wannier}), hence justifying our single-band per layer approximation.
Fig.~\ref{SFig1_wannier} displays both the DFT bands (black lines), as well as the tight-binding bands from the Wannier projection. The full-$d$ projection is displayed in red and the $d_{x^2-y^2}$-only one in blue, respectively. The energy windows of DFT bands for Wannier projections are set as: -1.0\,eV to 3.0\,eV, -3.0\,eV to 3.0\,eV, -1.5\,eV to 2.5\,eV and -3.0\,eV to 8.0\,eV for Nd$_3$Ni$_2$O$_6$ $d_{x^2-y^2}$ only,  Nd$_3$Ni$_2$O$_6$ Ni-$d$, Nd$_6$Ni$_5$O$_{12}$ $d_{x^2-y^2}$ only, and Nd$_6$Ni$_5$O$_{12}$ Nd-$d$+Ni-$d$ projections.
The agreement between DFT and Wannier bands is remarkable for both $d_{x^2-y^2}$, Ni-$d$ and Nd-$d$+Ni-$d$ projections, even for bands with a van-Hove singularity around the Fermi energy.

% ----------------------------------------------------------
\section{3. Orbital occupation} 
\label{sec:orbital_occ}
% ----------------------------------------------------------

With the exception of the undoped bilayer system  Nd$_{3}$Ni$_{2}$O$_{6}$ where also the Ni 3$d_{xz/yz}$ bands form a quasiparticle peak at the Fermi energy, all nickelates considered only have one strongly correlated band crossing the Fermi energy:
the Ni 3$d_{x^2-y^2}$ band. In some cases there are additional pockets (bands) that are however only very weakly renormalized compared to the DFT bands. They also do not show a strong broadening nor any formation of (Hubbard) side peaks.
This strongly suggests that strong correlations --and thus most likely the mechanism for high-temperature superconductivity-- resides in these effective Ni 3$d_{x^2-y^2}$ band. If pockets are present as e.g.~for the infinite layer nickelate system, they do not hybridize with this Ni 3$d_{x^2-y^2}$  band. All those reasons indicate that the minimal model for superconductivity is the effective Ni 3$d_{x^2-y^2}$ band with an appropriate filling.

In the multi-orbital Hamiltonian, this {\em effective} single Ni 3$d_{x^2-y^2}$ band is made up  out of predominately Ni 3$d_{x^2-y^2}$ character with some admixtures of the other Ni and Nd orbitals, due to hybridization effects. If we would include the oxygen $p$ orbitals in the multi-orbital calculation they would also contribute to this band. 
Hybridization with orbitals lying in energy above the Ni 3$d_{x^2-y^2}$ orbital 
such as the Nd states will decrease the occupation of the 3$d_{x^2-y^2}$ orbital
in the multi-orbital calculation, while hybridization with orbitals lying below such as the other Ni orbitals (or oxygen orbitals if included) will enhance the occupation of the Ni 3$d_{x^2-y^2}$ orbital.
If we consider a single-band effective low-energy Hamiltonian, such hybridization effects must not be included since it would give the wrong occupation of the {\em effective} orbital and the wrong Fermi surface. 

If all the other bands are completely below or above the Fermi energy, they are fully occupied or fully empty, and the
occupation of the effective low-energy Hamiltonian is simply given by the electrons remaining after filling all orbitals below the Fermi energy.
If there are, however, pockets present, which do not hybridize noticeable with the target band and are hence not included in the low-energy effective Hamiltonian, their occupation has to be accounted for. 

For transition metal oxides, this is the established DFT+DMFT procedure in case of a projection onto only the five $d$-orbitals or onto three $t_{2g}$ or two $e_g$  orbitals \cite{kotliar2004strongly,held2007electronic}. In our case, the  hybridization of the  Ni 3$d_{x^2-y^2}$ with the other Ni and Nd orbitals is rather weak so that the difference of the occupation of the  3$d_{x^2-y^2}$ orbital in the calculation with all orbitals as employed in \cite{Kitatani2020} and the  occupation of the {\em effective} Ni 3$d_{x^2-y^2}$ with all other orbitals filled or empty (except for pockets) is of only a few percent. 

In the following  we explicitly describe how we calculated the occupation of the  {\em effective}  single orbital Hamiltonian (Ni 3$d_{x^2-y^2}$ orbital)
using the latter procedure without the aforementioned hybridization effects.

In Table~\ref{table:occupation} we list both the occupation as well as the quasi-particle weight as obtain within DFT and DFT+DMFT for the full calculation with all orbitals. Hybridization is clearly present within the Ni $d$-shell, as illustrated by the occupation of the Ni $d_{z^2}$ orbital. This hybridization, is not a consequence of correlations, but already present within DFT. For a comparison, we further list the layer dependent occupation for the $d_{x^2-y^2}$-only model in Table~\ref{table:dx2y2_only_occupation} for different average occupations. 
 
In the case, where pockets are present which do not hybridize with the correlated Ni $d_{x^2-y^2}$ and hence are not included in the D$\Gamma$A calculation, the extraction of the correct filling becomes more involved. Below we outline the corresponding approach used for all compounds: 
 
\subsubsection*{Occupation Nd$_{6}$Ni$_{5}$O$_{12}$}

Once local correlations are taken into account the Nd derived pockets are shifted above the Fermi surface and can be considered as empty. Thus there is an average filling of $0.8$ electrons per Ni layer for the effective  $d_{x^2-y^2}$  orbitals in DMFT. For DFT, on the other hand, there is a pocket. Here, the occupation of the Ni $d_{x^2-y^2}$ orbitals has been obtained by adjusting the chemical potential of the $d_{x^2-y^2}$-only projection to lie on top of the full-$d$ projection, and subsequently integrating the spectral function from $-10$\,eV to $0$\,eV. Due to the presence of Nd derived pockets within DFT, this occupation is lower than $0.8$ on average.

\subsubsection*{Occupation Sr$_{x}$Nd$_{1-x}$NiO$_2$}

For the infinite layer case, we performed a Wannier projection onto the Ni $d_{x^2-y^2}$, Nd $d_{xy}$ and Nd $d_{z^2}$ orbital, which do not hybridize notably. We then fix the position of the pocket bands to their respective location within the DMFT calculation and again integrate the resulting spectral function of the two pockets from $-10$\,eV to $0$\,eV to obtain the number of electrons in the pockets $n_{\rm pockets}$. The occupation of the Ni $d_{x^2-y^2}$ orbital is then given by $1-x-n_{\rm pockets}$. This is the only case, where we have to employ an approximation, namely that the shape of the pocket and its spectral weight do not change by electronic correlations. For the pocket this is indeed a good 
approximation. We also cross-checked our result by computing the DMFT Fermi-surface and using Luttinger's sum rule \cite{Luttinger1960,LuttingerWard1960,Heath2020} to obtain the particle density.

\subsubsection*{Occupation  La$_{2+\delta}$Zr$_{1-\delta}$Ni$_2$O$_6$}
  
For $\delta=0.4$ the pockets are pushed above the Fermi surface by local correlations within DMFT. Hence the corresponding filling is simply given by the nominal valence 3$d^{9-\delta/2}$ (3$d^{8.8}$ when $\delta=0.4$) and correspondingly the occupation of Ni $d_{x^2-y^2}$ is $n=0.8$.

\begin{table*}
\caption{Full-$d$ DMFT and DFT orbital-resolved occupation, renormalization factor $Z$, and effective mass  $m^*$/$m_b$ of Ni-$d_{x^2-y^2}$ Wannier orbital for  the fully fledged calculation with all 3$d$-orbitals  (and for the pentalyer and infinite layer Nd 5$d$ orbitals).}
\label{table:occupation}
\begin{tabular}{ c|c|c|c|c|c|c|c|c}
\hline
\hline
System  & Atom & n($d_{xy}$) & n($d_{yz/xz}$) & n($d_{x^2-y^2}$) & n($d_{z^2}$) & n$_{total}$ & $Z$(Ni-$d_{x^2-y^2}$) & $m^*$/$m_b$ \\
  \hline
NdNiO$_2$  & Nd & 0.252 & 0.009 & 0.012 & 0.101 &   0.385 & \\
 (DMFT)    &  Ni & 1.961 & 1.927 & 0.945 & 1.854& 8.615 & 0.227 & 4.404\\
\hline
Nd$_{0.8}$Sr$_{0.2}$NiO$_2$  & Nd & 0.247 & 0.009 & 0.012 & 0.083 & 0.362 & &  \\
 (DMFT)    &  Ni & 1.957 & 1.917 & 0.822 & 1.824 & 8.438 & 0.355 & 2.812 \\
\hline
Nd$_3$Ni$_2$O$_6$ (n$_d$=8.5 in DMFT)    &  Ni & 1.994 & 1.850 & 0.824 & 1.980 & 8.500 & 0.348 & 2.867 \\
\hline
Nd$_3$Ni$_2$O$_6$ (n$_d$=8.5 in DFT)    &  Ni & 2.000 & 1.995 & 0.534 & 1.977  & 8.500 & & \\
\hline
Nd$_3$Ni$_2$O$_6$ (n$_d$=8.8 in DMFT)    &  Ni & 1.995 & 1.990 & 0.836 & 1.987 & 8.800 & 0.327 & 3.061 \\
\hline
Nd$_3$Ni$_2$O$_6$ (n$_d$=8.8 in DFT)    &  Ni & 2.000 & 1.997 & 0.814 & 1.992 & 8.800 & \\
\hline
Nd$_6$Ni$_5$O$_{12}$   & Nd-1 & 0.212 & 0.009 & 0.011 & 0.067 & 0.310 & & \\
 (DMFT)                & Nd-2 & 0.248 & 0.016 & 0.010 & 0.083 & 0.375 & & \\
    & Nd-3 & 0.041 & 0.011 & 0.003 & 0.016 & 0.083 & & \\
    & Ni-1 & 1.965 & 1.929 & 0.828 & 1.835 & 8.488 & 0.351 & 2.844 \\
    & Ni-2 & 1.966 & 1.929 & 0.821 & 1.849 & 8.496 & 0.352 & 2.838 \\
    & Ni-3 & 1.966 & 1.929 & 0.848 & 1.823 & 8.497 & 0.333 & 2.997 \\
\hline
    Nd$_6$Ni$_5$O$_{12}$   & Nd-1 & 0.275 & 0.012 & 0.015 & 0.095 & 0.410 & &  \\
 (DFT)  & Nd-2 & 0.315 & 0.022 & 0.014 & 0.114 & 0.487 & &  \\
    & Nd-3 & 0.066 & 0.016 & 0.004 & 0.025 & 0.128 & & \\
    & Ni-1 & 1.958 & 1.921 & 0.773 & 1.805 & 8.380 & & \\
    & Ni-2 & 1.960 & 1.919 & 0.762 & 1.828  & 8.388 & & \\
    & Ni-3 & 1.960  & 1.920 & 0.806 & 1.789 & 8.395 & & \\
\hline
\hline
\end{tabular}
\label{Tab:HoppingElements}
\end{table*}

% In Tab.~2 of the main text we showed the occupation for systems with a different number of layers for both DFT and DMFT. Here we report the layer-dependent occupation and quasiparticle weight $Z$ for the Ni $d_{x^2-y^2}$ only projection as described in Section~\ref{sec:eff_mass}. We vary the average occupation via a static doping (i.e. band-structure shift). Layer-dependent occupation is displayed in Tab.~\ref{table:dx2y2_only_occupation} and $Z$ in Tab.~\ref{table:dx2y2_only_Z}. A average filling per layer of $n = 0.83$ yields the best approximation to the Ni $d_{x^2-y^2}$ occupation of the full d-shell projection as displayed in the main text. 

\begin{table}[ht]
\caption{DMFT resulted occupation of the Ni 3 $d_{x^2-y^2}$ orbital of Nd$_6$Ni$_5$O$_{12}$, for a projection on only the Ni-$d_{x^2-y^2}$ band.}
\label{table:dx2y2_only_occupation}
\begin{tabular}{ c|c|c|c|c|c|c|c|c|c}
\hline
\hline
average occupation & 0.75  & 0.8   & 0.82  & 0.83  & 0.84  & 0.85  & 0.9   & 0.95  & 1    \\ \hline
Ni-1 (center-layer)& 0.746 & 0.801 & 0.817 & 0.829 & 0.841 & 0.851 & 0.9   & 0.95  & 1.004 \\ \hline
Ni-2 (interface)& 0.739 & 0.791 & 0.812 & 0.823 & 0.832 & 0.843 & 0.891 & 0.941 & 0.994 \\ \hline
Ni-3 (subinterface)& 0.764 & 0.808 & 0.827 & 0.838 & 0.847 & 0.857 & 0.91  & 0.958 & 1.004 \\ \hline
\hline
\end{tabular}
\end{table}

% ----------------------------------------------------------
\section{4. Effective mass} 
\label{sec:effmass}
% ----------------------------------------------------------

\begin{figure}[H]
\centering
\includegraphics[width=0.8\textwidth]{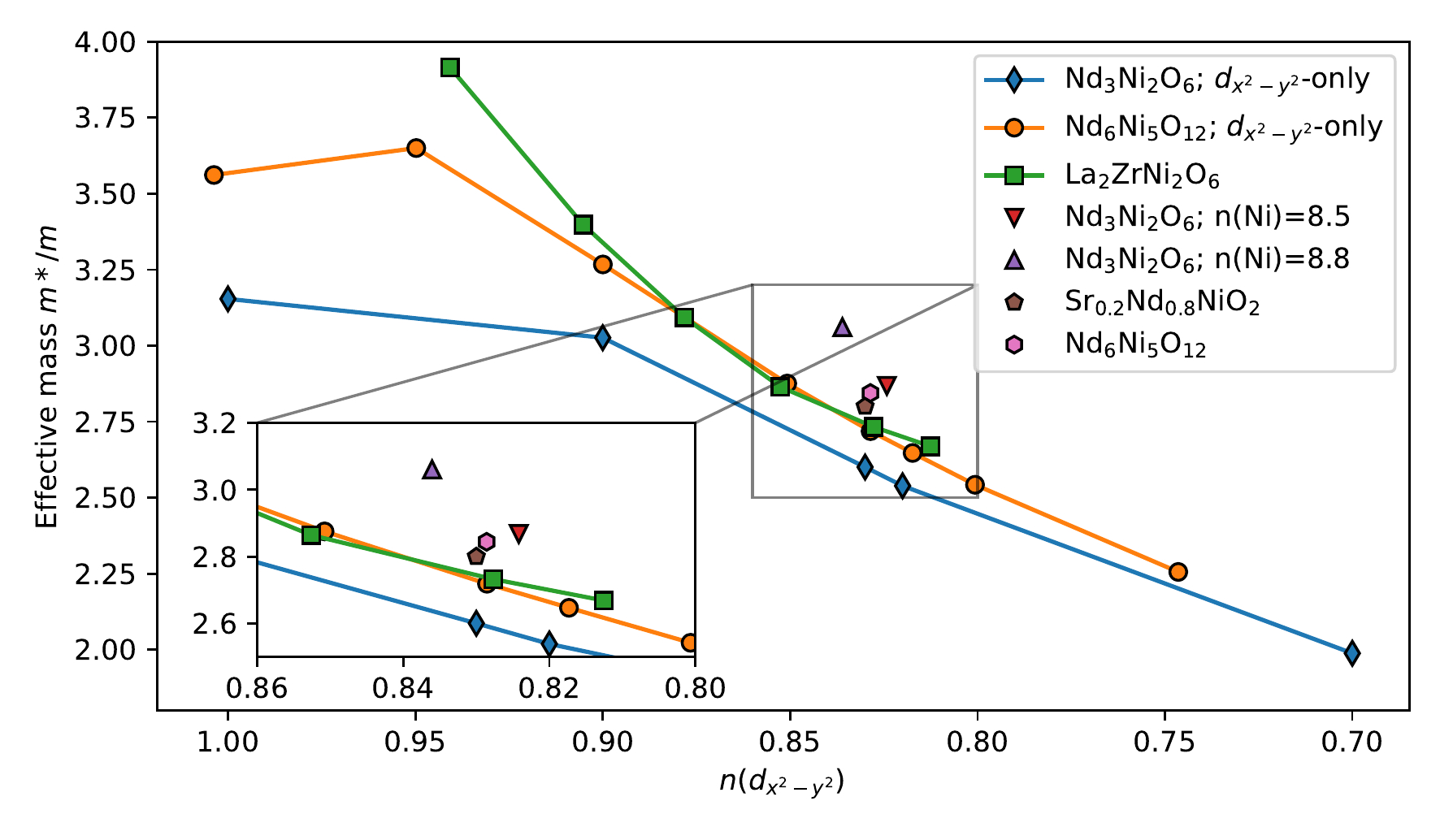}
\caption{Effective mass $m^*/m_b$ as a function of the occupation of the Ni $d_{x^2-y^2}$ orbital for all systems considered. For comparison we also included the values for the infinite layer nickelate from Ref.~\cite{Kitatani2020}.}
\label{SFig2_Z}
\end{figure}

DMFT allows us to compute the mass renormalization created by local correlations. As discussed in the main text, the quasi-particle weight $Z$ and thus also the effective mass $m^*/m_b = 1/Z$ are stunningly similar for the infinite layer and the pentalayer compound. We display this effective mass $m^*/m_b$ as a function of the occupation in the Ni $d_{x^2-y^2}$ orbital in Fig.\ref{SFig2_Z} for all systems considered in the main text and the strontium doped infinite layer nickelate from \cite{Kitatani2020}. In case of the pentalayer system we display the values for the layer at the center, but the ratio between $m^*/m_b$ and occupation is almost the same for all layers. We find it quite remarkable, that both known superconducting nickelates (the doped infinite layer and the quintuple layer) exhibit almost the same effective mass $m^*/m_b\sim2.85$ at the same Ni 3$d_{x^2-y^2}$ occupation. La$_{2+x}$Zr$_{1-x}$Ni$_{2}$O$_6$ shows both qualitatively and quantitatively a similar mass renormalization as the superconducting compounds. We believe this to be one indication for the possibility of superconductivity in La$_{2+x}$Zr$_{1-x}$Ni$_{2}$O$_6$. 

Let us also emphasize that the mass renormalization in the fully fledged calculation with all $d$-orbitals in Table~\ref{Tab:HoppingElements} are quite similar to that in the projection onto a single Ni 3$d_{x^2-y^2}$ band in Table~\ref{table:dx2y2_only_Z}.

%The downturn of the pentalayer single-orbital $m^*/m$ at half-filling  in Fig.\ref{SFig2_Z} is %most likely a numerical artifact. %within the numerical accuracy.

% Hence, for future theoretical material design the effective mass can be another indicator to asses the possibility of superconductivity in a certain compound \pw{maybe I am "leaning out of the window" here}. Nevertheless, it can clearly only be taken in the case of a one band Fermi surface, as can be seen from the undoped Nd$_{3}$Ni$_{2}$O$_{6}$ ($n=8.5$) system. Here the effective mass $m^*/m ~ 2.87$  and the filling $n = 0.824$ is close to the superconducting structures, but the appearance of the $d_{xz/yz}$ orbital at the Fermi surface will destroys the one-band picture and possibly superconductivity. 

\begin{table}[ht]
\caption{DMFT resulted quasi-particle renormalization $Z$ of the $d_{x^2-y^2}$ orbital of Nd$_6$Ni$_5$O$_{12}$, for a projection on only the Ni-$d_{x^2-y^2}$ band.}
\label{table:dx2y2_only_Z}
\begin{tabular}{ c|c|c|c|c|c|c|c|c|c  }
\hline
\hline
 average occupation   & 0.75  & 0.8   & 0.82  & 0.83  & 0.84  & 0.85  & 0.9   & 0.95  & 1    \\ \hline
Ni-1 (center-layer) &0.443 & 0.393 & 0.378 & 0.368 & 0.356 & 0.348 & 0.306 & 0.274 & 0.281\\ \hline
Ni-2 (interface) & 0.452 & 0.402 & 0.383 & 0.373 & 0.365 & 0.355 & 0.314 & 0.281 & 0.281 \\ \hline
Ni-3 (subinterface) & 0.427 & 0.385 & 0.367 & 0.357 & 0.349 & 0.34  & 0.296 & 0.266 & 0.275 \\ \hline
\hline
\end{tabular}
\end{table}

% ----------------------------------------------------------
\section{5. DMFT spectral function} 
\label{sec:d_only_projection}
% ----------------------------------------------------------

In the main text we presented the $k$-resolved (integrated) spectral function $A(k,\omega)$ ($A(\omega)$) for the bilayer in Fig.~\ref{Fig4} and the pentalayer compound in Fig.~\ref{Fig5}. This section supplements these results, by also providing said quantities for (i) the bilayer system at a different filling ($n$(Ni)=8.8) to simulate electron doping effect, (ii) the bilayer system, but for the $d_{x^2-y^2}$ only projection at $n$=0.8 and (iii) the quintuple layer system but for the $d_{x^2-y^2}$ only projection.

\begin{figure}[ht]
\includegraphics[width=\textwidth]{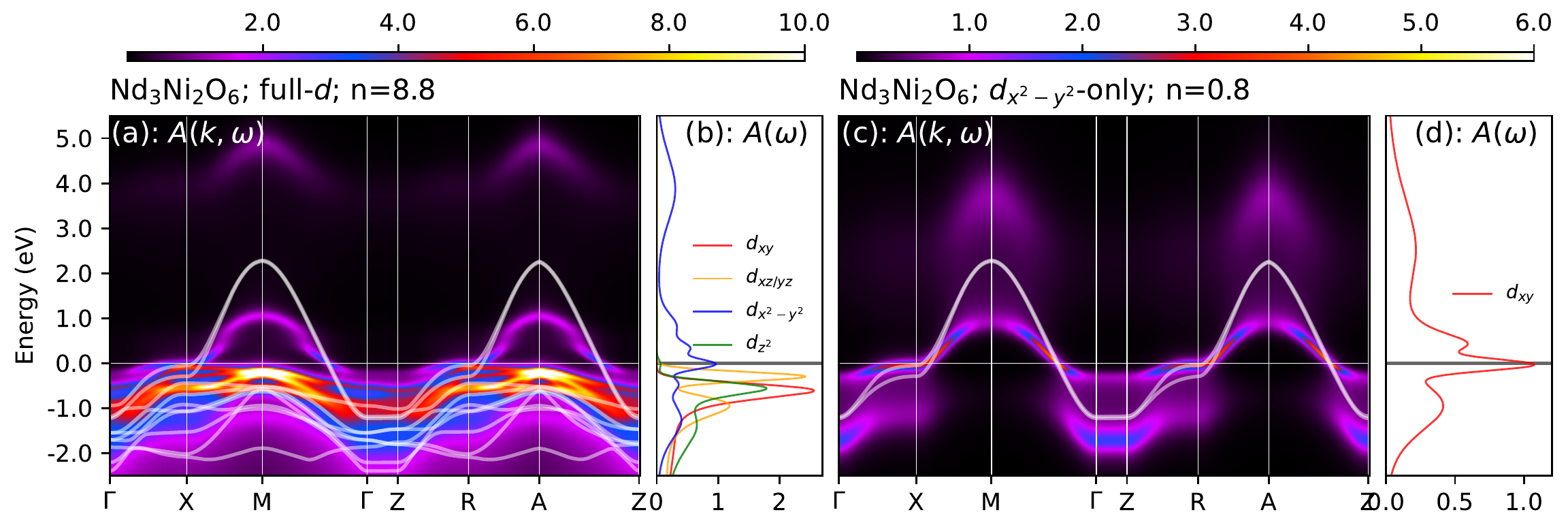}
\caption{(a) DMFT $k$-resolved spectral functions $A(k,\omega)$ of Nd$_3$Ni$_2$O$_6$. (b) $k$-integrate spectral function $A(\omega)$. Solid lines in (a) are DFT Wannier-bands. The average occupation per nickel site is $n=$8.8. The $d_{xz/yz}$ orbital is below the Fermi energy in contrast to $n=$8.5 from the main text.}
\label{SFig:Awk_N2}
\end{figure}

Information (i) is displayed in Fig.\ref{SFig:Awk_N2}(a)-(b) and shows two interesting features compared to the undoped ($n$(Ni)=8.5) case as discussed in the main text: First, the $d_{xz/yz}$ orbital is fully occupied and pushed below the Fermi surface, hence recovering the one-band physics. Second, the quasipartilce weight $Z = 0.326$ is comparable to the superconducting compounds as displayed in Fig.~\ref{SFig2_Z}. Thus we conclude, that the doped bilayer Nd$_3$Ni$_2$O$_6$ compound falls much more into the paradigm of known nickelate superconductors. In section \ref{sec:La2ZrNi2O6} we will propose La$_{2+x}$Zr$_{1-x}$Ni$_{2}$O$_6$ as promising candidate compound to achieve such a doping in a bilayer system. 

%Second, because most of the electrons go into the $d_{xz/yz}$ orbital the filling of the most relevant $d_{x^2-y^2}$ orbital is almost unchanged. It evolves from $n(d_{x^2-y^2}) = 0.824$ at $n(Ni) = 8.5$ to $n(d_{x^2-y^2}) = 0.836$ at $n(Ni) = 8.8$. A similar statement is true about the quasi particle weight, which evolves from $m^*/m = 0.348$ in the undoped case to $m^*/m = 0.326$. Thus we conclude, that the doped bilayer Nd$_3$Ni$_2$O$_6$ compound falls much more into the paradigm of known nickelate superconductors. In section \pw{ref correct section} we will propose one promising candidate compound to achieve such a doping in a bilayer system. 

Information (ii) is displayed in Fig.\ref{SFig:Awk_N2}(c)-(d) for completeness.

\begin{figure}[ht]
\includegraphics[width=\textwidth]{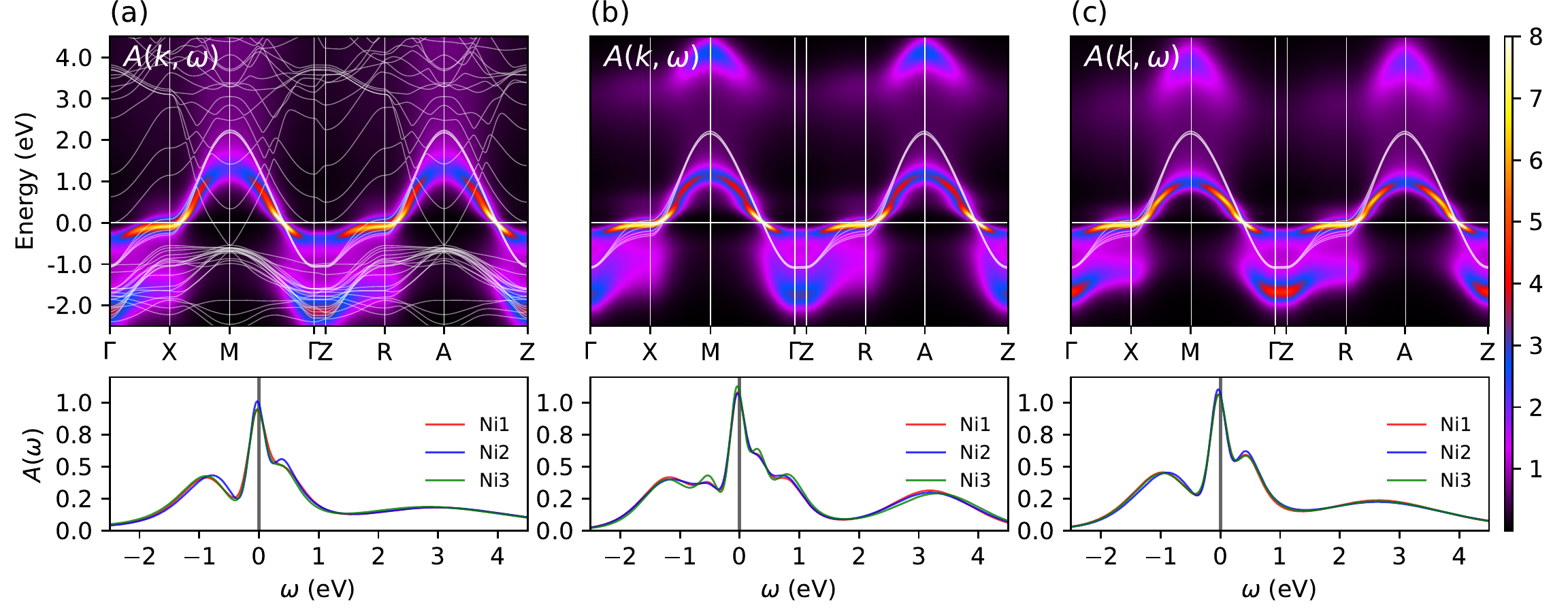}
\caption{DMFT $k$-resolved spectral functions $A(k,\omega)$ of Nd$_6$Ni$_5$O$_{12}$  (top) and $k$-integral $A(\omega)$ (bottom). (a) Ni $d_{x^2-y^2}$ orbital from full-$d$ projection. (b) {\em effective} Ni $d_{x^2-y^2}$ orbital from $d_{x^2-y^2}$-only projection. (c) same as (b), but with a larger $\alpha$ for the analytic continuation.
}
\label{SFig:Awk_N5}
\end{figure}

Information (iii) is displayed in Fig.~\ref{SFig:Awk_N5}(b) and highlight how DMFT suppresses layer dependent physics. Not only do local correlations lead to a more uniform occupation, also the $k$-integrated spectral function $A(\omega)$ is quite similar for all layers in Fig.~\ref{SFig:Awk_N5}(a). The most notable differences is (i) that the upper Hubbard band is more spread out in the full-$d$ projection and (ii) that there is additional structure in the low-frequency part of the $d_{x^2-y^2}$-only projection. Both differences can partially be attributed to the analytic continuation. We used more statistic for the $d_{x^2-y^2}$-only projection since it contains fewer bands and is thus numerically less demanding. In turn there is less noise and more features can be resolved by the maximum entropy method. To compare this we show analytic continuation for the $d_{x^2-y^2}$-only projection, but with a higher $\alpha$ in the MaxEnt code for the $d_{x^2-y^2}$-only projection in Fig.\ref{SFig:Awk_N5}(c), which leads  broader features and better agreement with the calculation including all five Ni orbitals. Nevertheless, while the position of the upper Hubbard band peak depends on the MaxEnt parameters, it's weight and center of mass does not in a significant way. The corresponding spectral weights $w_{uhb}$ and center of masses $c_{uhb}$ are $\{w_{uhb} =1.48, c_{uhb} = 3.3\}$ for the $d_{x^2-y^2}$-only and $\{w_{uhb} =1.55, c_{uhb} = 4.0\}$ for the full-$d$ projection.

% ----------------------------------------------------------
\section{6. Fermi surface} 
\label{sec:fs}
\begin{figure}[ht]
\includegraphics[width=\textwidth]{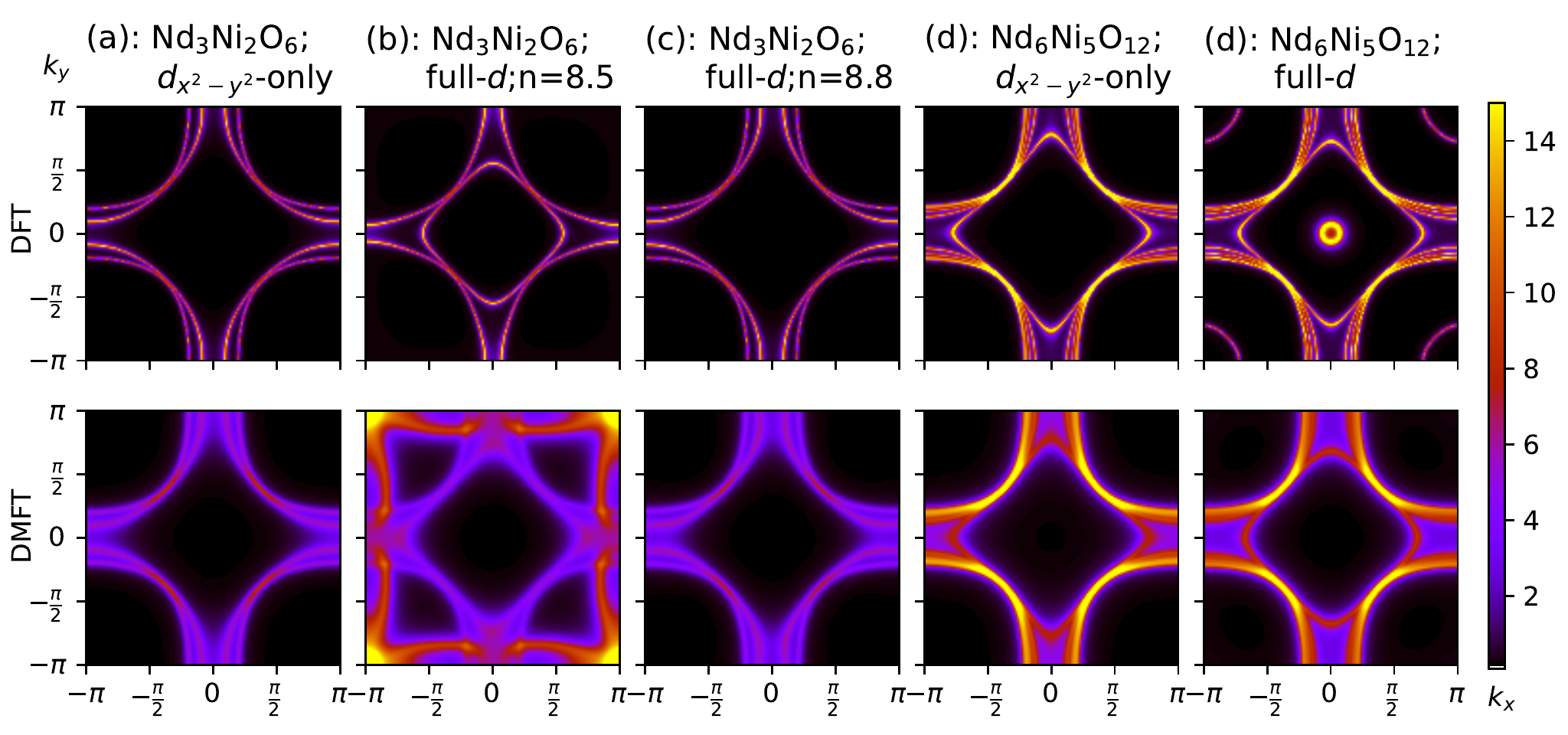}
\caption{Fermi surface for the DFT solution (top) and for DFT+DMFT (bottom). (a) Nd$_{3}$Ni$_{2}$O$_{6}$ for the $d_{x^2-y^2}$-only projection and average occupation per layer $n=0.8$. (b) Nd$_{3}$Ni$_{2}$O$_{6}$ for the full-$d$ projection and a total of $8.5$ electrons. (c) same as (b) but for $8.8$ electrons. (d) Nd$_{6}$Ni$_{5}$O$_{12}$ for the $d_{x^2-y^2}$-only projection and average occupation per layer $n=0.8$. (d) Nd$_{6}$Ni$_{5}$O$_{12}$ for the full-$d$ projection. Notice how the pockets are shifted above the Fermi surface once local correlations are taken into account in (d) [same as Fig.~\ref{Fig4} (e,f) of the main paper, re-plotted here for better comparison].}
\label{SFig:FS_N2_N5}
\end{figure}

\begin{figure}[ht]
\includegraphics[width=0.7\textwidth]{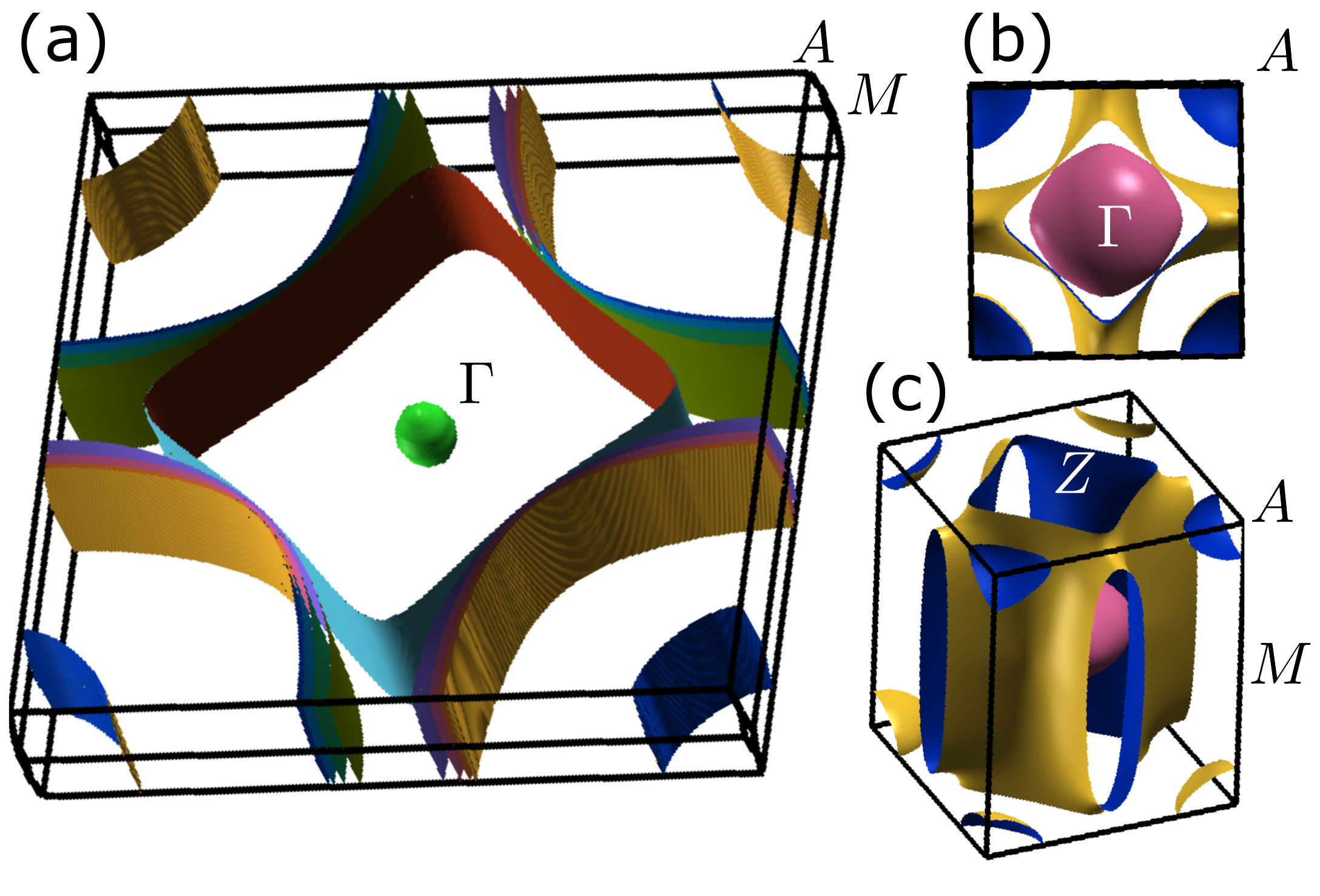}
\caption{(a) Nd$_{6}$Ni$_{5}$O$_{12}$ Fermi surface for the DFT solution within the 1st BZ. The pocket at the $\Gamma$-point and the small tube around the $A$ and $M$ point are Nd-$d$ derived, while the five large sheets derive from the Ni $d_{x^2-y^2}$ orbitals. (b) NdNiO$_2$ DFT Fermi surface top view and (c) side view. Here the pocket at $\Gamma$-point is much more pronounced and the second pocket is only around the $A$-point.  }
\label{DFT_Wannier_FS}
\end{figure}

% ----------------------------------------------------------
Since superconductivity is a low-temperature phenomenon it is most strongly influenced, if not dominated by, low-energy excitations. Hence, a lot of the system can be learned by looking at the Fermi surface, or, for finite temperatures, at the spectral function around zero energy. For simplicity we will refer to both concept just as Fermi surface. 
Fig.~\ref{SFig:FS_N2_N5}(d) displays the Fermi surface for the Nd$_{6}$Ni$_{5}$O$_{12}$ system within the DFT framework (top) and DFT+DMFT (bottom). The DFT solution still has pockets from the Nd atoms around the $\Gamma=[0,0,0]$ and a tube-like pocket extending around the $M=[\pi,\pi,0]$ to the $A=[\pi,\pi,\pi]$ point. To better visualize these pockets (tubes) we also plotted the DFT Fermi-surface in a 3D fashion in Fig.~\ref{DFT_Wannier_FS}(a) using the Xcrysden program package \cite{KOKALJ1999} together with the DFT Fermi-surface of NdNiO$_2$ (b)-(c).  These are, however, pushed above the Fermi surface once local correlations are included with the DMFT framework. Another point worth mentioning is the appearance of one electron-like Fermi surface and four hole-like ones, which is due to one of the five Wannier bands of the pentalayer  having its von-Hove singularity  above the Fermi-energy (see also discussion in Section \ref{sec:dft_wannier}). However, there exists no assignment of these different Fermi surfaces to a single layer; each layer contains a superposition of them all.

When discussing Nd$_{3}$Ni$_{2}$O$_{6}$ in the main text (Fig.~\ref{Fig3} in main) one feature emerging from local correlations was the $d_{xz/yz}$ orbital contribution to the Fermi surface predominantly at the $M$-point. Here in Fig.~\ref{SFig:FS_N2_N5}(b) we display this effect by showing the systems Fermi surface. Notice, that in DFT (top) the holes reside in the $d_{x^2-y^2}$ orbital only. The  recovery of a $d_{x^2-y^2}$ only Fermi surface via electron doping to $n$(Ni)=8.8 is displayed in Fig.~\ref{SFig:FS_N2_N5}(c).

In the main text we estimated $T_c$ from an effective one-band model, whose Wannier projection was discussed in Section \ref{sec:dft_wannier}. Here we show the corresponding Fermi surface for the Nd$_{6}$Ni$_{5}$O$_{12}$ and the Nd$_{3}$Ni$_{2}$O$_{6}$ system at an average $d_{x^2-y^2}$ occupation of $0.80$ in Fig.~\ref{SFig:FS_N2_N5}(d) and Fig.~\ref{SFig:FS_N2_N5}(a), respectively. 

% ----------------------------------------------------------
\section{7. The bilayer nickelate {La$_{2}$ZrNi$_2$O$_6$}} 
\label{sec:La2ZrNi2O6}
% ----------------------------------------------------------

\begin{figure}[ht]
\includegraphics[width=\textwidth]{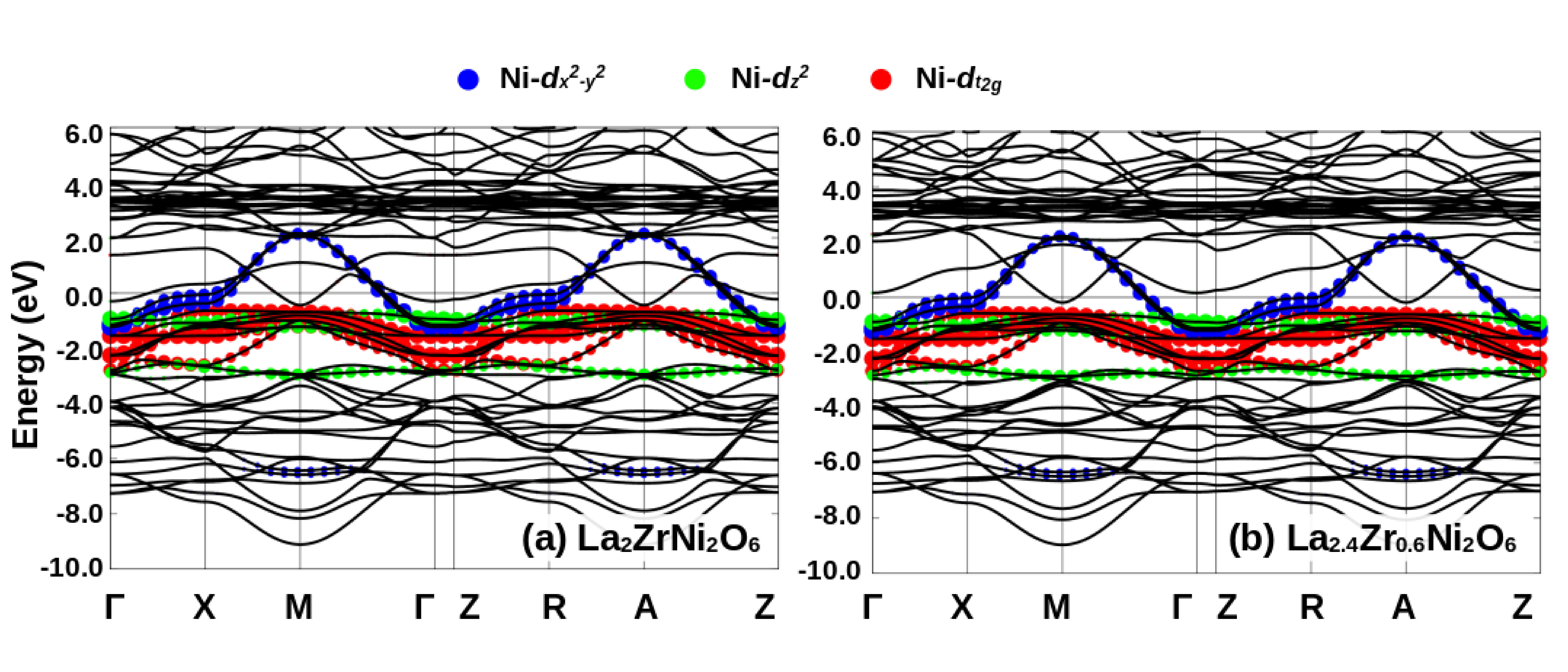}
\caption{DFT bands (black lines) and orbital character as indicated by color. (a) La$_{2}$ZrNi$_2$O$_6$ and (b) La$_{2.4}$Zr$_{0.6}$Ni$_2$O$_6$}
\label{SFig10.5_La_DFT_Wannier}
\end{figure}

\begin{figure}[ht]
\includegraphics[width=\textwidth]{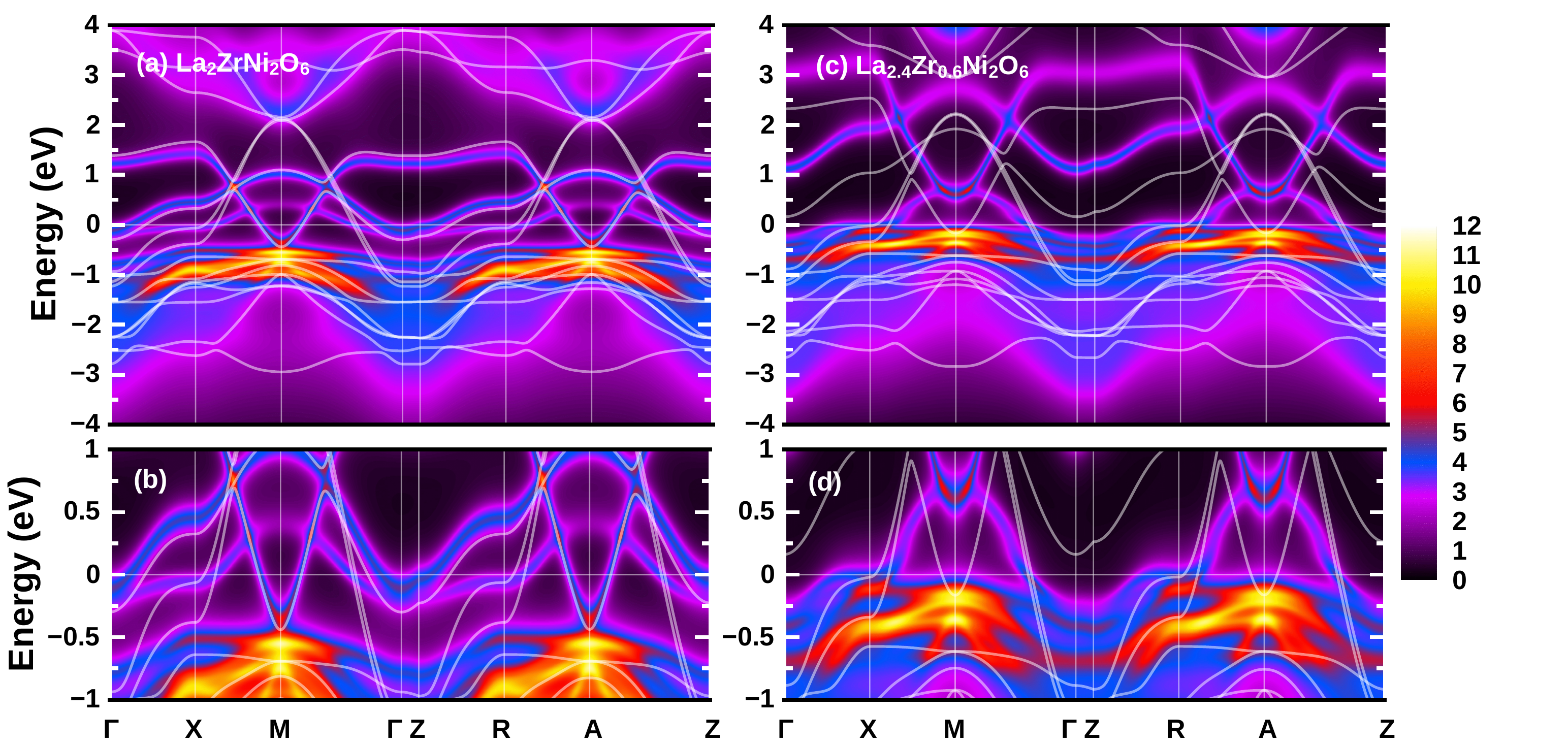}
\caption{DMFT resulted $k$-resolved spectral function $A(\omega,k)$ for La$_{2}$ZrNi$_2$O$_6$ (a) and La$_{2.4}$Zr$_{0.6}$Ni$_2$O$_6$ (c). White lines correspond to the DFT Wannier bands. Bottom displays a zoom-in around the Fermi energy of the corresponding plot in the top. }
\label{SFig11_La_Awk}
\end{figure}

\begin{figure}[ht]
\includegraphics[]{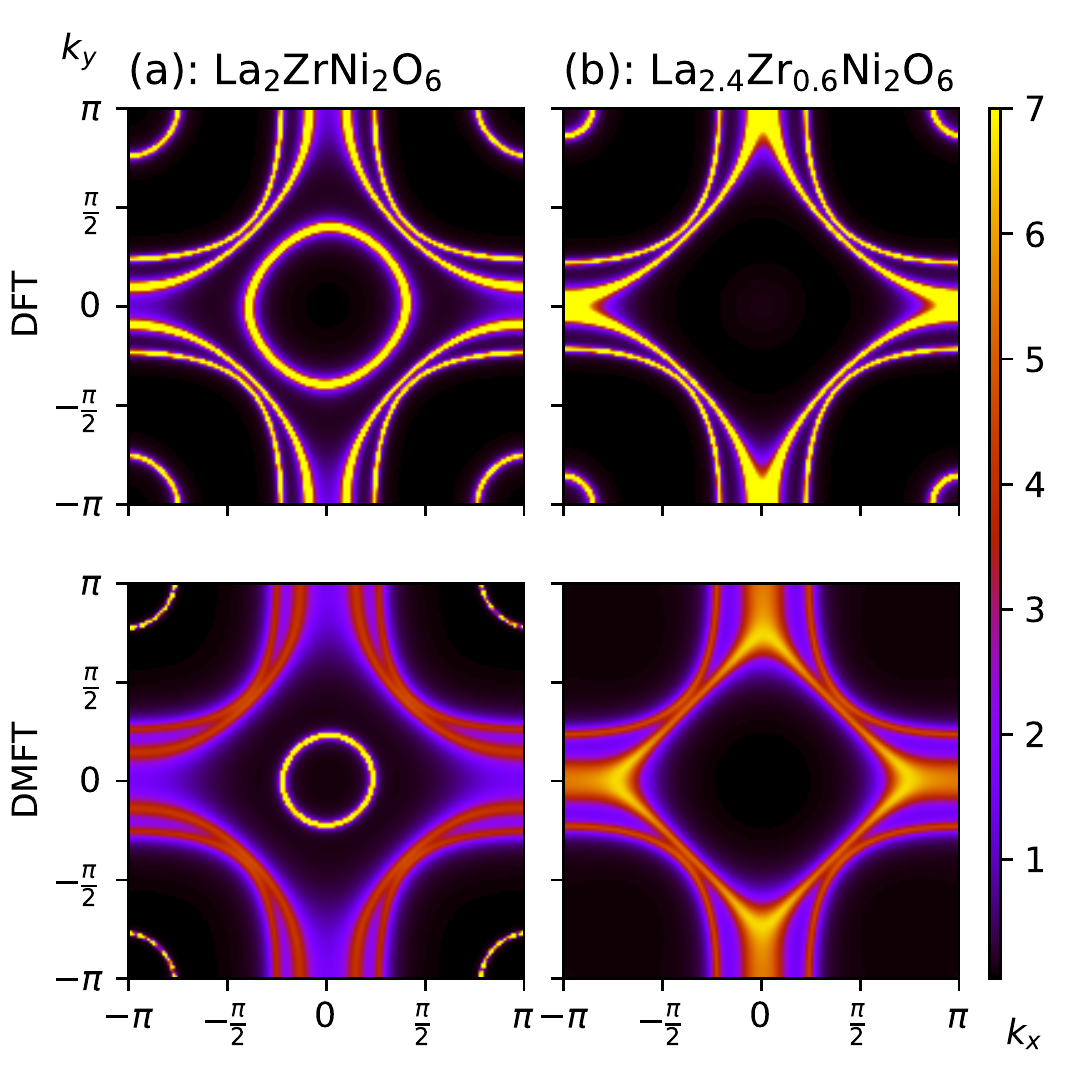}
\caption{Fermi surface for (a) La$_{2}$ZrNi$_2$O$_6$ and (b) La$_{2.4}$Zr$_{0.6}$Ni$_2$O$_6$. Top displays DFT Wannier and bottom DFT+DMFT. }
\label{SFig:LaZrNiO_FS}
\end{figure}

In the main text we discussed the possibility to engineer a superconducting bilayer nickelate compound. Given the results of Fig.~\ref{SFig:LaZrNiO_FS}(b), one promising candidate for a nickelate superconductor  is La$_{2+\delta}$Zr$_{1-\delta}$Ni$_2$O$_6$ for $\delta=0.4$. In Fig.~\ref{SFig10.5_La_DFT_Wannier} we display the orbital character of the Ni $d$-shell Wannier projection for the parent compound ($100\%$ Zr doping: left) and at $40\%$ Zr doping (right). Wannier projection of the Zr $d$-shell is not shown there for clarity, but instead as white lines in Fig.~\ref{SFig11_La_Awk} on top of the DFT+DMFT $k$-resolved spectral function. For $40\%$ Zr doping the disappearance of the pockets due to local correlations is clearly shown in Fig.~\ref{SFig11_La_Awk}(c,d). This is further emphasised by the Fermi surface plots in Fig.~\ref{SFig:LaZrNiO_FS}(a) for the undoped and Fig.~\ref{SFig:LaZrNiO_FS}(b) for the doped compound. 

% Additionally, we show the $k$-resolved spectral function in Fig.~\ref{SFig11_La_Awk} and the Fermi surface for $\delta=0.0$ and $\delta=0.4$ in Fig.~\ref{SFig:LaZrNiO_FS}(a) and Fig.~\ref{SFig:LaZrNiO_FS}(b), respectively. 

\end{document}